\documentclass[prl,nofootinbib,twocolumn]{revtex4}
\pdfoutput=1
\usepackage{graphicx}

\usepackage{amsmath} 
\usepackage{amssymb}
\usepackage{bm}

\def\be{\begin{equation}}
\def\ee{\end{equation}}
\def\ba{\begin{eqnarray}}
\def\ea{\end{eqnarray}}

\def\f{\frac}
\def\l{\left}
\def\r{\right}

\begin{document}
\title{\large \bf  Scalar radiation from Chameleon-shielded regions}
\author{Alessandra Silvestri}
\affiliation{Department of Physics, MIT, Cambridge, Massachusetts 02139, USA}
\begin{abstract}I study the profile of the chameleon field around a radially pulsating mass. Focusing on the case in which the background (static) chameleon profile exhibits a thin shell, I add small perturbations to the source in the form of time-dependent radial pulsations. It is found that the chameleon field inherits a time dependence and there is a resultant scalar radiation from the region of the source. This has several interesting and potentially testable consequences.
\end{abstract}

\maketitle

Chameleons are scalar fields characterized by a profile that depends on the local  matter density, as a consequence of their coupling to matter fields. Such a coupling naturally arises in string and supergravity theories and is often encountered in models  addressing the phenomenon of cosmic acceleration. For instance, several models of dark energy introduce a coupling in the dark sector~\cite{Copeland:2006wr}, while modified theories of gravity typically introduce a nonminimally coupled scalar degree of freedom~\cite{Silvestri:2009hh}. In all these cases the scalar has a nonlinear potential and its effective mass  depends on the local matter density, i.e. it displays a chameleon behavior~\cite{Khoury:2003aq}. It is precisely this nonlinear mechanism that allows the field to have a nontrivial dynamics on cosmological scales, while escaping stringent constraints from local tests of gravity even if it couples to matter with gravitational strength. This screening mechanism has been suggested in~\cite{Khoury:2003aq} and further explored in~\cite{Waterhouse:2006wv,Tamaki:2008mf,Hu:2007nk}. 

As analyzed in~\cite{Satz:2004hf,Moffat,Dai:2008zza}, Birkhoff's theorem ceases to  hold in these models and there are several interesting consequences of this fact. An intriguing and yet unexplored feature of these theories is that the coupling between the chameleon and matter fields allows a radially pulsating mass to directly propagate a disturbance into the space surrounding it. Interestingly, this would have several consequences which in principle are testable. On the astrophysical side, compact sources could emit scalar radiation and the rate of energy loss from a binary black hole would be bigger than in general relativity. Also, the dynamics of core collapse might be significantly modified in the presence of a chameleon field. Moreover,  in conformally coupled theories, the particle masses and the coupling constants depend on the chameleon field and, if the profile of the field is time-dependent, these quantities will all inherit a time dependence. 

These are all interesting scenarios which require a detailed analysis of the chameleon profile around a  time-dependent source. In this Letter I specialize to the case of an isolated radially pulsating mass. I start with the time-independent case and  determine the profile of the chameleon around a static spherically symmetric mass following~\cite{Khoury:2003aq,Tamaki:2008mf}.  Successively, specializing to those cases in which the chameleon profile displays a thin shell, I add a time-dependent perturbation to the source mass and study the corresponding time- and scale-dependent perturbation to the static profile. 

Let us consider theories described by the action
\be\label{Chameleon_action}
S=\int d^4x \sqrt{-g}\l[\tfrac{M_P^2}{2}R-\tfrac{1}{2}(\partial\phi)^2-V(\phi)\r]+S_m[\psi^{(i)}_m,g_{\mu\nu}^{(i)}]
\ee
where $\phi$ is the chameleon field, $\psi_m^{(i)}$ are matter fields and
\be\label{conf_metric}
g_{\mu\nu}^{(i)}=e^{2\beta_i\phi/M_P}g_{\mu\nu}
\ee
is the metric describing the geodesics for the $i$-th matter field. $\beta_i$ are dimensionless coupling constants, they can be assumed to be of order one and in general they may be different for each matter species. 

Varying action~(\ref{Chameleon_action}) with respect to the scalar field one obtains its equation of motion 
\be\label{Ch_eq}
\Box\phi=V_{,\phi}+\tfrac{\beta_i}{M_P} e^{4\beta_i\phi/M_P}g_{(i)}^{\mu\nu}T^{(i)}_{\mu\nu}
\ee
For nonrelativistic matter the trace of the energy-momentum tensor reduces to the energy density, $g_{(i)}^{\mu\nu}T^{(i)}_{\mu\nu}=\tilde{\rho}^{(i)}$; we shall use a conserved energy density defined as $\rho^{(i)}\equiv e^{3\beta_i\phi/M_P}\tilde{\rho}^{(i)}$. Concentrating on a single matter component (therefore dropping the $i$ indices), we can identify the right-hand side of~(\ref{Ch_eq}) with the following effective potential
\be\label{V_eff}
V^{\textrm{eff}}(\phi)\equiv V(\phi)+\rho e^{\beta\phi/M_P}
\ee
i.e. the field profile depends on the local matter distribution. 

\underline{\emph{Background profile}}
Let us start from the static case, solving for the chameleon profile in the presence of an isolated static spherically symmetric mass of density $\rho_c$ and radius $R_c$, immersed in a background of homogeneous density $\rho_{\rm G}$. We shall work in the weak gravity regime, assuming that the Newtonian potential is small everywhere and the backreaction of the energy density in the field $\phi$ is negligible. Therefore the metric can be approximated to the one of a Minkowski spacetime, i.e. $g_{\mu\nu}\approx\eta_{\mu\nu}$. This static scenario has been studied in~\cite{Khoury:2003aq,Waterhouse:2006wv,Tamaki:2008mf}. The energy density coupled to the chameleon is
\be\label{rhobackground}
\rho_0(r)=\l\{\begin{array}{cc}\rho_c,& \,\,\,r< R_c\\\rho_{\rm G},&\,\,\,r>R_c\end{array}\r.
\ee
We consider inverse power-law potentials of the form
\be\label{Ch_pot}
V(\phi)=\f{M^{4+n}}{\phi^n}
\ee
where $M$  has units of mass and $n$ is a positive constant, $n\geq 1$. The runaway nature of the potential is important for the chameleon mechanism~\cite{Khoury:2003aq} and the inverse power-law realization (\ref{Ch_pot}) arises in supersymmetric models as well as from nonperturbative effects in string theory and is desirable in quintessence models~\cite{Steinhardt:1999nw,Copeland:2006wr}. It is also the form of a potential characteristic of $f(R)$ theories in the Einstein frame. In what follows I use the approximation $\beta\phi/M_P \ll 1$ which holds through the expansion history as shown in~\cite{Khoury:2003aq,Waterhouse:2006wv}.

For a given value of the background density, $\rho_i$, the mass around the corresponding minimum of the effective potential, $\phi_i$, has the form
\be\label{m}
m^2_i\equiv n(n+1)\f{M^{4+n}}{\phi_{i}^{n+2}}+\f{\beta^2\rho_i}{M_P^2}
\ee
I will indicate with $(\phi_c, m_c)$ and $(\phi_{\rm G}, m_{\rm G})$  the value of the field and the mass at the minimum of the effective potential respectively for $\rho_i=\rho_c$ and $\rho_i=\rho_{\rm G}$. 

The full problem can be solved numerically, however with few approximations one can easily obtain an analytical solution~\cite{Khoury:2003aq,Waterhouse:2006wv,Tamaki:2008mf}. For the purpose of this Letter it is not relevant to review the details of the solution, but simply notice that depending on the parameters of the model there are different types of profiles and it will suffice to separate them in two classes, as follows.  Assuming that the field will eventually settle into the external minimum, $\phi_{\rm G}$, far enough from the source, the field can remain close to $\phi_{\rm G}$ also inside the source or, if the perturbing effect of the source is strong enough,  the field can be driven to the internal minimum $\phi_c$ inside the source. It depends on whether the potential energy cost for not lying at the minimum inside the source outweighs the gradient energy gain for not introducing a profile in the field. In other words,  whether the source is strong enough to perturb the field profile depends on the relative magnitude of  $\phi_{\rm G}-\phi_c$ and $6\beta M_P\Phi_c$, where $\Phi_c$ is the Newtonian potential at the surface of the source. If $(\phi_{\rm G}-\phi_c)/(6\beta M_P\Phi_c)\ll1$, the field remains close to $\phi_{\rm G}$ outside the source,  close to the minimum $\phi_c$ inside the source and interpolates between the two minima over a thin radial region $\Delta R_c$ close to the surface of the mass. This solution is known as the {\it thin-shell} configuration and more details about its form and the conditions under which it is achieved can be found in~\cite{Khoury:2003aq,Waterhouse:2006wv,Tamaki:2008mf,Tsujikawa:2009yf}; a complete classification of the different configurations can be found in~\cite{Waterhouse:2006wv,Tamaki:2008mf}.

The thin-shell configuration is the one of interest in cosmology since the scalar field can have a nontrivial dynamics on cosmological scales, while the fifth force that it mediates on local scales is suppressed by the thin-shell coefficient
\be\label{thin-shell-coeff}
\f{\Delta R_c}{R_c}\equiv\f{\phi_{\rm G}-\phi_c}{6\beta M_P\Phi_c}\ll 1
\ee 
This way, the chameleon field can easily hide from local tests of gravity while, for instance, sourcing the cosmic acceleration on large scales~\cite{Khoury:2003aq}.

\begin{figure}[t]
\includegraphics[width=1\columnwidth]{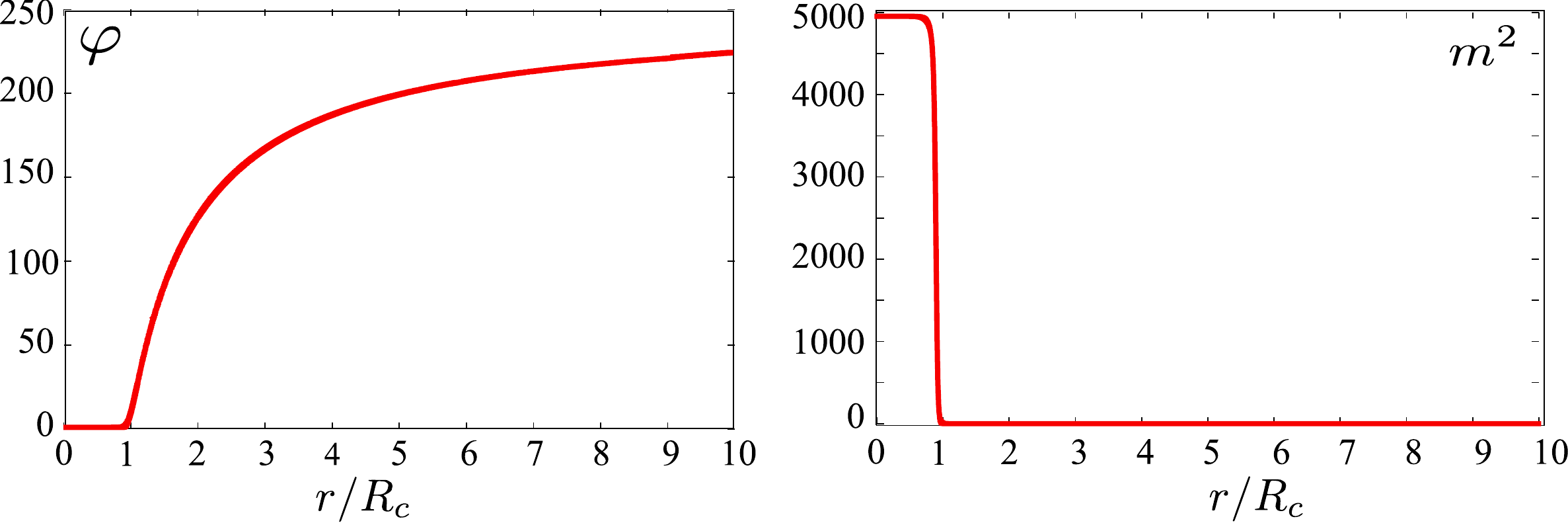}
\caption{Chameleon profile $\varphi$ ({\it left}) and mass squared ({\it right}) in presence of an isolated static spherically symmetric source of radius $R_c$. The parameters of the model are: $n=1$, $\rho_{\rm G}/\rho_c=2\cdot 10^{-5}$ and $(m_cR_c)^{-1}=0.01420777$ as in~\cite{Tsujikawa:2009yf}.}
\label{Ch_back}
\end{figure}
I will use the static profile as background for the time-dependent analysis, therefore it is desirable to obtain a precise solution by numerical methods, treating the problem as a boundary value problem, with boundary conditions
\be\label{BCs_back}
\f{d\phi}{dr}(r=0)=0\,\,,\,\,\,
\phi(r\rightarrow\infty)=\phi_{\rm G}
\ee
I use the {\it relaxation} algorithm~\cite{NR} and rescale the field and coordinates defining  $\varphi\equiv \phi/\phi_c$ and $x\equiv r/R_c$. The equation for the chameleon reads
\be\label{num_background}
\f{d^2\varphi}{dx^2}+\f{2}{x}\f{d\varphi}{dx}=\f{(m_cR_c)^2}{n+1}\l[\f{\rho_0(x)}{\rho_c}-\f{1}{\varphi^{n+1}}\r]\,,
\ee
where I have used $\f{\beta^2\rho_c}{M_P^2}\ll1$~\cite{Khoury:2003aq}. For appropriate choices of the parameters it is possible to recover the thin-shell configuration. I restrict to these cases and show a representative solution for the profile and the mass of the chameleon in Fig.\ref{Ch_back}. One can notice a sharp transition in the field and mass values over a thin shell; otherwise the values are, to good approximation, constant with $(\varphi,m)\approx(1,m_c)$ in the inner region ($x< 1$) and $(\varphi,m)\approx(\varphi_G,m_G)$ in the outer region ($x>1$).

\underline{\emph{Time-dependent profile}}
Let us now add small, time-dependent perturbations to the source mass. In particular, let us specialize to the case of radial pulsations and implement them keeping the mass of the source fixed and considering oscillations of its surface around the radius $R_c$. Including the time dependence, the chameleon now obeys the following equation
\be\label{time-eq_dimless}
-\f{\partial^2\varphi}{\partial \tau^2}+\f{\partial^2\varphi}{\partial x^2}+\f{2}{x}\f{\partial\varphi}{\partial x}=\f{(m_cR_c)^2}{n+1}\l[\f{\rho(x)}{\rho_c}-\f{1}{\varphi^{n+1}}\r]\,,
\ee
where $\tau\equiv t/R_c$ is a dimensionless time.

We shall decompose the field and the density in a background static component and a time- and scale-dependent perturbation, respectively $\varphi(\tau,x)=\varphi_0(x)+\delta\varphi(\tau,x)$ and  $\rho(\tau,x)\equiv\rho_0(x)[1+\delta(\tau,x)]$; $\varphi_0(x)$ satisfies Eq. (\ref{num_background}). The perturbation $\delta\varphi$ will  obey the equation
\ba\label{time-eq-num}
-\f{\partial^2\delta\varphi}{\partial \tau^2}+\f{\partial^2\delta\varphi}{\partial x^2}+\f{2}{x}\f{\partial\delta\varphi}{\partial x}=\,\hspace{3cm}&&\nonumber\\
\f{(m_cR_c)^2}{n+1}\l[\varphi_0^{-(n+1)}-\l(\varphi_0+\delta\varphi\r)^{-(n+1)}+\f{\rho_0}{\rho_c}\,\delta\r]&&
\ea
\begin{figure}[t]
\includegraphics[width=0.65\columnwidth]{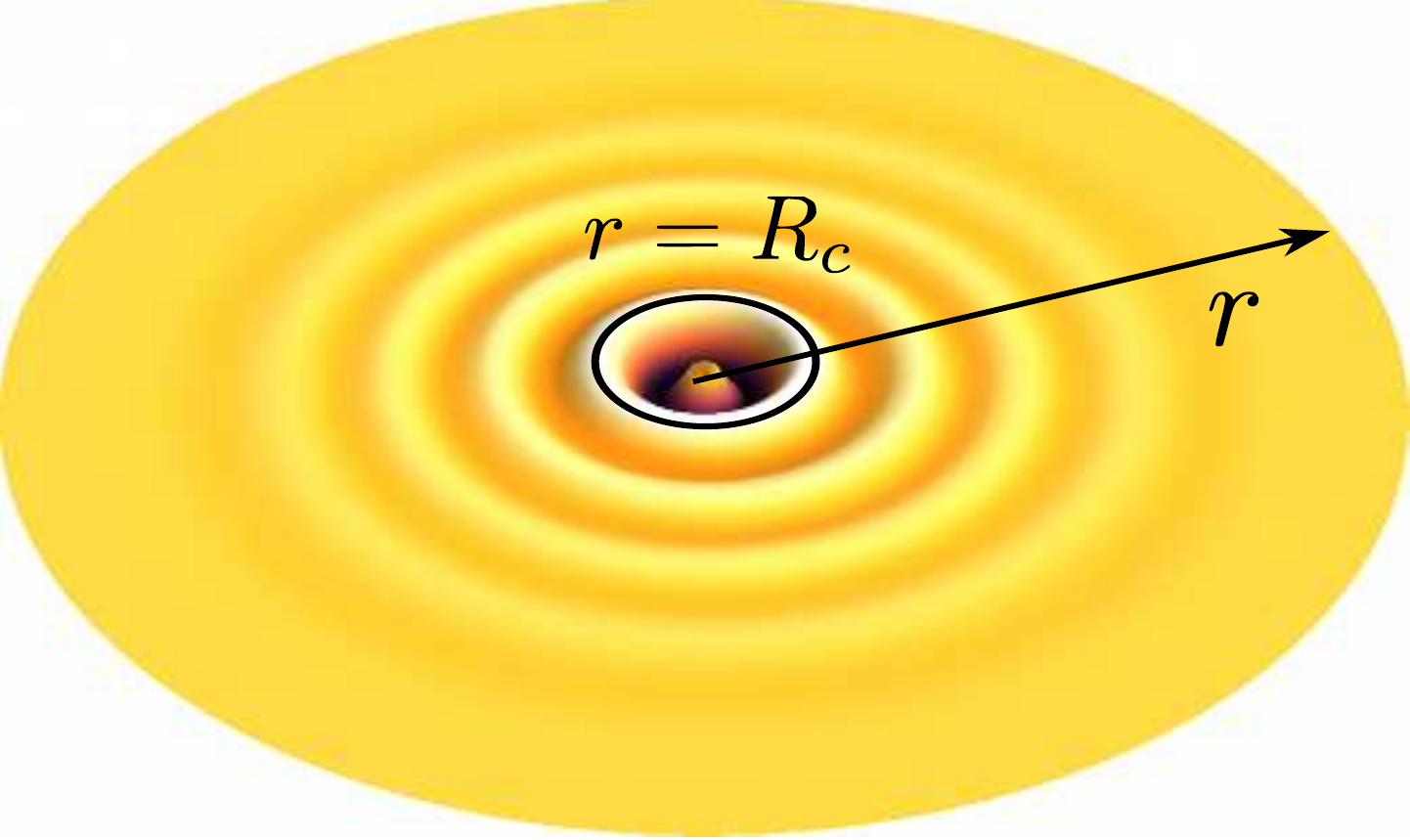}
\caption{Time-dependent perturbation to the static chameleon profile as a function of radial distance from the source center on an equatorial section at time $t=13 R_c$ for the model~(\ref{model})}
\label{Ch_osc_3D}
\end{figure}
Linearizing in $\delta\varphi$ one gets 
\be\label{time-eq-num_linear}
-\f{\partial^2\delta\varphi}{\partial \tau^2}+\f{\partial^2\delta\varphi}{\partial x^2}+\f{2}{x}\f{\partial\delta\varphi}{\partial x}=\f{(m_cR_c)^2}{\varphi_0^{n+2}}\delta\varphi+\f{(m_cR_c)^2}{n+1}\,\delta\,.
\ee

This is an inhomogeneous Klein-Gordon equation with a scale-dependent mass given by the mass of the background field $\varphi_0$. 
We shall solve it for $\tau>0$ with the following initial and boundary conditions
\be\label{ICBCs}
\delta\varphi(0,r)=0=\partial_t\delta\varphi(0,x)\,\,\,,\,\,\,
\lim_{x \to (0,\infty)}\delta\varphi(\tau,x)= 0\,.
\ee

As stated above, we hold fixed the mass and geometry of the spherical source, and vary its volume via small oscillations of the surface. In other words, we assume that its radius oscillates around $R_c$ with frequency $\omega_0$, i.e. $R_c[1+\delta R_c/R_c\sin{(\omega_0 R_c \tau)}]$ with $\delta R_c/R_c\ll1$, therefore the density perturbation can be written as
\be\label{source}
\delta(\tau,x)=3\f{\delta R_c}{R_c}\sin(\omega_0 R_c \tau)\theta(\tau)\theta(1-x)
\ee
where $\theta(y)$ is the Heaviside step function.

\begin{figure}[t]
\includegraphics[width=0.8\columnwidth]{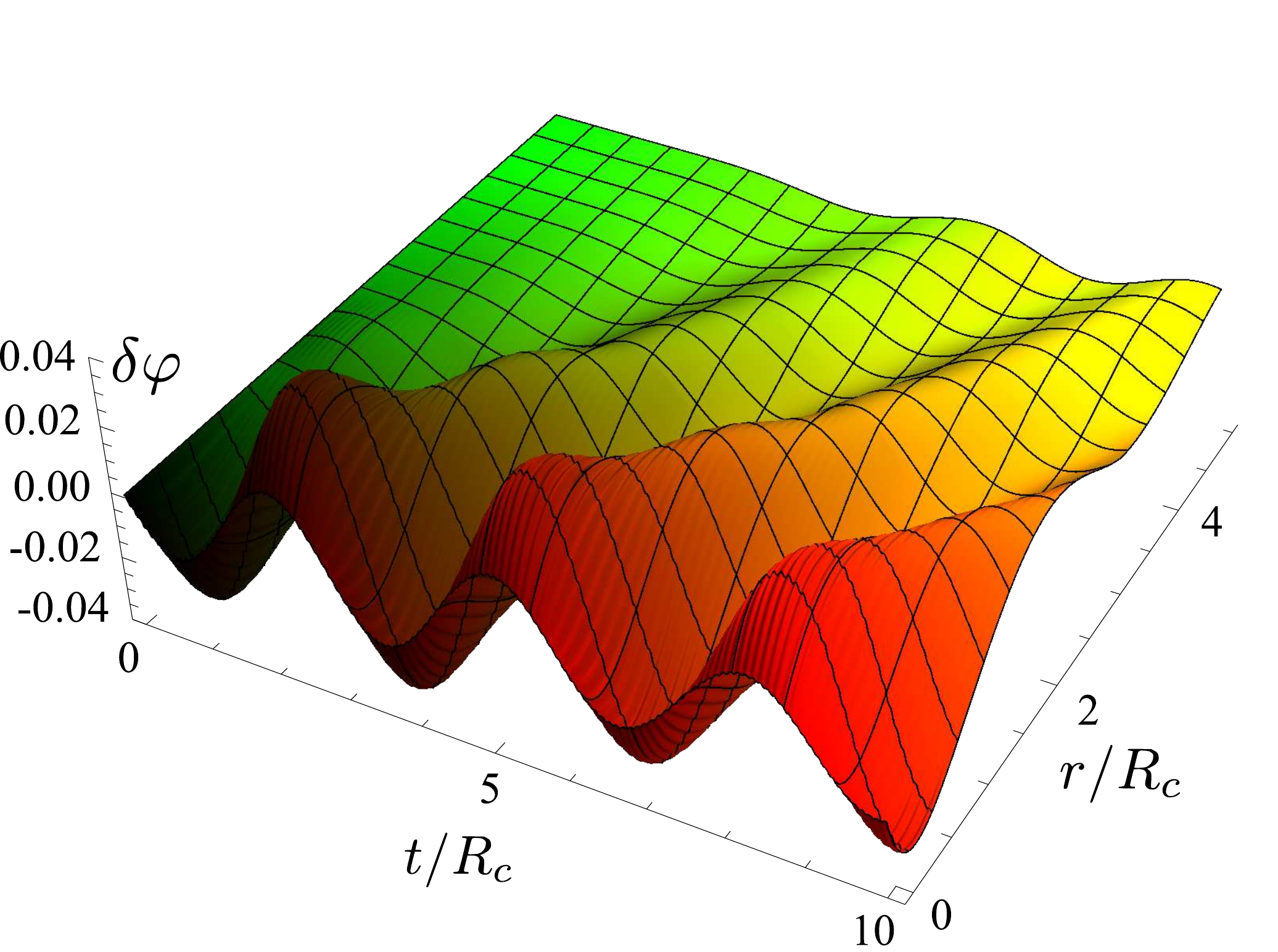}
\caption{Time-dependent perturbation $\delta \varphi$ to the static chameleon profile as a function of time and radial distance from the source center for the model~(\ref{model}).}
\label{Ch_osc_1}
\end{figure}

\begin{figure}[t]
\includegraphics[width=1\columnwidth]{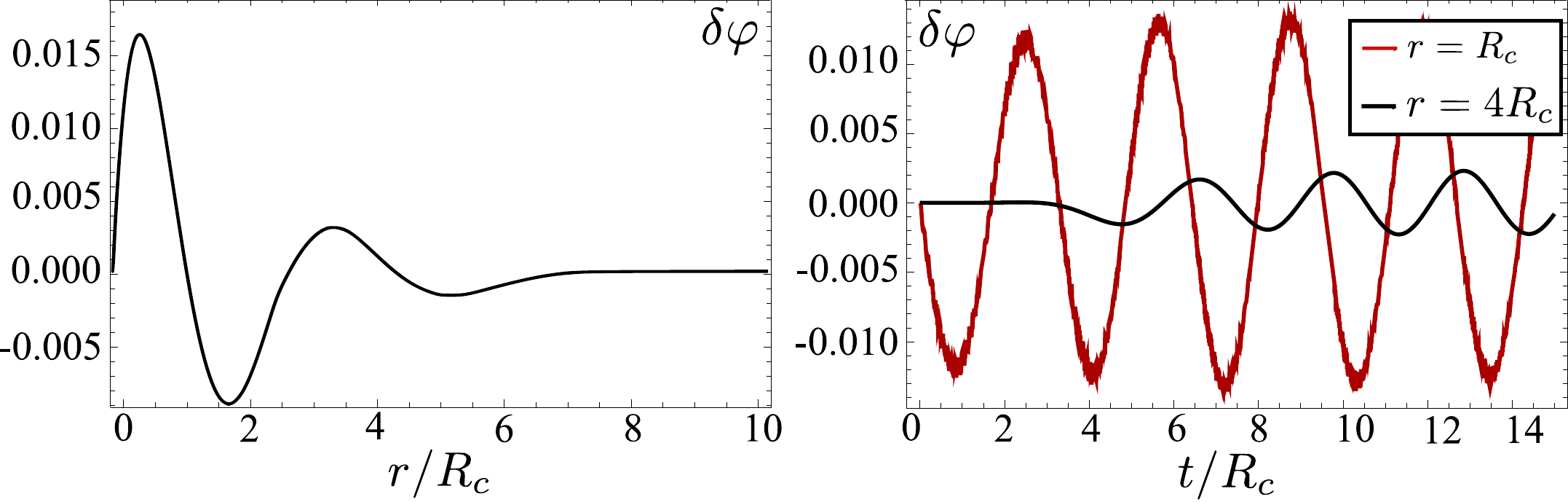}
\caption{{\it Left}: Perturbation to the profile of the chameleon as a function of the radial distance at time $t/R_c=5$. {\it Right}: Perturbations to the profile of the chameleon as a function of time at the distances $r/R_c=1$ (red thick curve) and $r/R_c=4$ (black thin curve). Both plots correspond to the model~(\ref{model}).}
\label{Ch_osc_2}
\end{figure}

I have solved numerically both the full (\ref{time-eq-num}) and  linear (\ref{time-eq-num_linear}) equation and found that, as long as $\delta\varphi/\varphi\ll1$, the linear solution is a very good approximation to the full one. Therefore in the remaining we shall focus on the linear equation (\ref{time-eq-num_linear}). I show some representative features of the solution in Figures \ref{Ch_osc_3D}, \ref{Ch_osc_1} and \ref{Ch_osc_2} for the model 
\ba\label{model}
&&n=1, \,\,\rho_{\rm G}/\rho_c=2\cdot 10^{-5},\,\, (m_cR_c)^{-1}=0.01420777,\nonumber\\ 
&&\omega_0 R_c=2,\,\,\delta R/R_c=0.01
\ea
 where I have used the numerical solution of Eq. (\ref{num_background}) for the background profile $\varphi_0$.
As can be noticed from Figures \ref{Ch_osc_3D} and \ref{Ch_osc_1}, the oscillations propagate outside the source in the form of ripples in the profile of the chameleon, i.e. there is an outgoing {\it scalar radiation}. Looking at Fig. \ref{Ch_osc_2} it can be noticed that the field  responds to the pulsations of the source differently in the region inside ($x<1$) and outside ($x>1$) the source. The amplitude of the oscillations in the region outside the source is smaller and the ratio of the amplitudes in the inner and outer region depends on the parameters of the model, in particular on  the ratio of the masses $m_c/m_{\rm G}$. Performing a Fourier analysis of the oscillations we find a superposition of two modes of oscillations, related to the frequency of the source $\omega_0$ and to the internal mass $m_c$.  In the right panel of Fig. \ref{Ch_osc_2} the two modes can be noticed in the inner region, while in the outer the mode $\omega_0$ dominates over the other.  

With few approximations we can find an analytical solution to Eq.~(\ref{time-eq-num_linear}). It is a Klein-Gordon equation with a space-dependent mass which can be well approximated by a step function, as evident in Fig.\ref{Ch_back} . Therefore we can separate the spatial interval in two regions with constant mass, (i.e.  $m\approx m_c$ for $x<1$ and  $m\approx m_{\rm G}$ for $x>1$), and solve for two standard Klein-Gordon equations. We then match the solutions at the boundary $x=1$. 
Let us perform a Fourier transform with respect to the time coordinate $t$ and solve the equation in the frequency space. Depending on the value of the frequency $\omega$, we have to solve a spherical or modified spherical Bessel equation, homogeneous in the outer region and inhomogeneous in the inner region. After imposing the boundary conditions and matching the solutions at $x=1$ we find the following expression for the Fourier transform of $\delta\varphi$
\begin{equation}\label{an_sol}
\delta\tilde{\varphi}=\left\{\begin{array}{rl}
\frac{S_L}{|\lambda_c|^2}\l[x\cosh(\lambda_cx)-\tfrac{\sinh(\lambda_cx)}{\lambda_c}\r]\f{e^{-\lambda_cx}}{x}&x<1\\
\frac{S_Le^{-\lambda_c}}{|\lambda_c|^2}\l[\cosh(\lambda_c)-\tfrac{\sinh(\lambda_c)}{\lambda_c}\r]\frac{e^{-\lambda_{\rm G} (x-1)}}{x}&x>1
\end{array}\r.
\end{equation}
where $\lambda_i\equiv \sqrt{(m_iR_c)^2-(\omega R_c)^2}$, $i=c,G$ is real or imaginary depending on the value of $\omega$. The source term is
\be\label{source}
S_L\equiv\f{(m_cR_c)^2}{n+1}\f{3\,\delta R_c}{R_c}\f{\omega_0 R_c}{(\omega_0 R_c)^2-(\omega R_c)^2}\,.
\ee
In deriving (\ref{an_sol}) I have approximated the mass with a step function. While a more precise solution would model also the transition from $m_c$ to $m_{\rm G}$ over a finite thin shell, this transition was properly taken into account in the numerical solution and I have found good agreement between (\ref{an_sol}) and  the Fourier transform of the numerical solution. We notice that there are two resonances at $\omega=\omega_0$ and $\omega=m_c$ as found in the numerical solution. I derived the equivalent of (\ref{an_sol}) for the case in which the chameleon does not form a thin shell and $\phi_0=\phi_{\rm G}$ $\forall\, r$.  The frequency spectrum at a given point in space has a different shape and the resonances are now at $\omega_0$ and $m_{\rm G}$; therefore the resonance at $m_c$ signals that the field is close to $\phi_c$ inside the source, i.e.  the screening mechanism is at work. 

The oscillations in the chameleon carry away energy from the source. We can estimate the rate of energy loss by integrating the flux of the chameleon over a sphere of large radius $r$ and dividing by the energy of the source. This quantity would be useful to assess the detectability of the radiation,  however the result depends significantly on the type of source and  model parameters; therefore, for a meaningful estimate one needs to model properly physical sources, perhaps extending the formalism to the relativistic regime characteristic of strong field systems. I intend to address this in future work, however a preliminary estimate of the energy loss rate (averaged over one period of the source) for model~(\ref{model}) gives $\tfrac{dE/dt}{E}\approx O(10^{-21})s^{-1}$.  This value is smaller than known rates for binary pulsars~\cite{Will:2005va}, although model~(\ref{model}) is not strictly comparable to binary pulsars.  Once the analysis is properly extended to strong field systems, the energy loss rate might get closer to the observational threshold, offering the opportunity to constrain chameleon theories with binary pulsars or core-collapsing objects.

\underline{\emph{Conclusions}}
I have analyzed the profile of the chameleon field around a  radially pulsating source. As expected,  the pulsations perturb the surrounding space inducing ripples in the chameleon field that propagate as scalar radiation from the source. The physical consequences that could give rise to observable features are several.
For instance, energy could be drained from binary black holes faster than in general relativity, the ripples in the chameleon profile would induce time-variations of masses and fundamental coupling constants and the hydrodynamics of compact objects, in particular the core collapse, could be significantly modified. Finally, it is possible that when the time dependence of the source cannot be treated as a small perturbation, the changes in the profile of the chameleon would destroy the thin shell of the background profile. All these features are potentially observable and could offer new ways of testing chameleon type theories, complementary to existing tests~\cite{Brax:2004qh,Steffen:2010ze}. 

In this Letter the analysis was restricted to the case of a radially pulsating isolated source, modeled as a homogeneous solid sphere with small oscillations of the surface and the calculations were carried out in the weak field limit. Although simplistic, this scenario is a good approximation to many models and allows us to identify potentially observable features. The results are promising and motivate further work in this direction, with a more complex modeling of the sources. In particular it would be useful to extend the formalism to relativistic cases and to nonspherical configurations. 

I wish to thank E.~Bertschinger, L.~Giomi, L.~Pogosian for useful discussions and their feedback on this work. 
I acknowledge   support by NSF under grant AST-0708501.

\end{document}